# The progress of neutron texture diffractometer at China Advanced Research Reactor

LI MeiJuan[1], LIU XiaoLong[1], LIU YunTao[1]*, TIAN GengFang[1], GAO JianBo[1], YU ZhouXiang[1], LI Yuqing[1], WU LiQi[1], YANG LinFeng[1], SUN Kai[1], WANG HongLi[1], CHEN DongFeng[1]*

*Department of Nuclear Physics, China Institute of Atomic Energy, Beijing, 102413, China*

**Abstract:** The first neutron texture diffractometer in China has been built at China Advanced Research Reactor due to the strong demands of texture measurement with neutrons from domestic user community. This neutron texture diffractometer has high neutron intensity, moderate resolution and is mainly applied to study the texture in the commonly used industrial materials and engineering components. In this paper, the design and characteristics of this instrument are described. The results for calibration with neutrons and quantitative texture analysis of Zr alloy plate are presented. The comparison of texture measurement among different neutron texture diffractometer of HIPPO at LANSCE, Kowari at ANSTO and neutron texture diffractometer at CARR illustrates the reliable performance of this texture diffractometer.

**PACS:**

## 1 Introduction

The texture of a polycrystalline material is the preferred orientation distribution of its crystallites, and can be described by the orientation distribution function (ODF) with respect to a macroscopic sample coordinate system [1]. Texture is an intrinsic feature of metals, ceramics, polymers and rocks [2], it influences not only many anisotropic physical, but also technological properties of polycrystalline materials [3], and is regarded as one of the essential parameters for a full microstructural characterization of polycrystalline materials.

Texture determination is usually based on the pole figure measurement by x ray, synchrotron radiation, electron back-scatter diffraction and neutron diffraction. Among these techniques, neutron diffraction has its own advantages due to the low absorption coefficient of neutrons [3, 4] (the penetration depth for most materials in neutron diffraction is a factor of $10^2$-$10^3$ larger than X-ray diffraction). 1) Large sample can be used, thus high accuracy and good statistics as well as bulk texture can be obtained; 2) Accurate determinations of texture can be achieved for coarse-grained, texture inhomgeneity, multiphase samples and for samples with small volume fraction of second phase; 3) The scattering factors of neutron change irregularly with atomic number, even with the various isotopes of the same element, hence the texture studies of light element phases, especially in the presence of other phases containing heavy elements, will be much easier. 4) Neutron can be scattered magnetically, this effect is being used for magnetic texture analyses; 5) More importantly, texture measurements for samples in some environments such as high temperature and loading can be carried out. Texture analysis by neutron diffraction has become a standard method to investigate bulk textures of different types of materials [5]. As a consequence neutron texture diffractometers have been built in different neutron scattering laboratories in the world.

However, like other neutron scattering experiments, pole figure measurement by neutrons requires an intense neutron source such as a nuclear reactor with at least medium neutron flux or a spallation neutron source. The China Advanced Research Reactor (CARR) at China Institute of Atomic Energy (CIAE) with maximum thermal neutron flux of about $8\times10^{14}$/sec·cm$^2$ [6-7] is well qualified for neutron texture measurements and other neutron scattering experiments. A number of different type neutron scattering instruments have now been or are being built around the reactor. The Neutron Texture Diffractometer (NTD) is one of the neutron instruments built around CARR at present stage, and is the first platform for neutron texture measurement in China. In order to make full use of this instrument by the domestic user's community, the design, characteristics and performance of the newly constructed NTD at CARR will be introduced in this paper.

## 2  Instrument design and characteristics

The NTD is located at beam tube H2-1 of CARR, the main components of it and the layout are shown in figure 1.

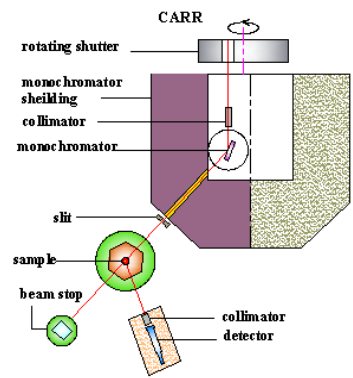

**Figure 1**   Layout of Neutron Texture Diffractometer at CARR

The design was performed based on the consideration of neutron intensity and the instrument resolution. Since pole figure measurements are very time-consuming, it is the most important factor for a texture diffractometer to have high neutron intensity. On the other hand, high resolution means that more well-separated reflection peaks can be chosen for pole figure measurements and samples with larger cells and more complicated structures can be investigated [8-9].

According to Caglioti, the instrument resolution function expressed by $A_{1/2}$, the full width at half height for the peaks, and the integrated peak intensity $L$ of the instrument are given as follows [10-12]:

$$A_{1/2}^2 = U \tan^2 \theta + V \tan \theta + W$$

$$U = \frac{4(\alpha_1^2 \alpha_2^2 + \alpha_1^2 \beta^2 + \alpha_2^2 \beta^2)}{\tan^2 \theta_M (\alpha_1^2 + \alpha_2^2 + 4\beta^2)},$$

$$V = \frac{-4\alpha_2^2(\alpha_1^2 + 2\beta^2)}{\tan \theta_M (\alpha_1^2 + \alpha_2^2 + 4\beta^2)},$$

$$W = \frac{\alpha_1^2 \alpha_2^2 + \alpha_1^2 \alpha_3^2 + \alpha_2^2 \alpha_3^2 + 4\beta^2(\alpha_2^2 + \alpha_3^2)}{\alpha_1^2 + \alpha_2^2 + 4\beta^2} \tag{1}$$

$$L = \frac{\alpha_1 \alpha_2 \alpha_3 \beta}{(\alpha_1^2 + \alpha_2^2 + 4\beta^2)^{1/2}} \tag{2}$$

where $\alpha_1$, $\alpha_2$ and $\alpha_3$ are the horizontal divergences of the primary, second and third collimators respectively, $\beta$ is the monochromator mosaic spread, $2\theta_M$ is the take-off angle of the monochromator.

As can be seen, the above two requirements are in conflict with each other. The instrument resolution can be easily improved by using collimators with smaller horizontal divergence. But at the same time, the diffractometer intensity will be decreased drastically. It is not possible for a texture diffractometer to achieve both the resolution and intensity as high as desired.

If the instrument resolution function is designed higher than that required for samples to be measured, a needless loss in diffraction intensity will be caused, and this is not permissible due to the fact that the reactor sources available in the world at present are not strong enough for neutron scattering experiments. Therefore a compromise has to be made between them. In the case of NTD at CARR, most of the materials to be measured on it are those with simple structures and relatively small cells. For the pole figure measurements of these materials, the instrument is required to have only moderate resolution, this enables it to achieve high neutron intensity. Obviously, In order to make the instrument reasonable and efficient, the best solution to this problem is to let the instrument resolution function just match the required one in a large enough $2\theta$ range.

To obtain the required resolution function, the expression for the separation between adjacent peaks for cubic cells deduced from the Bragg equation was appropriately used in the design, here, λ is the neutron wavelength, 2θ is the scattering angle and a is the size of the unit cell. It is clear that the required resolution function is symmetrical about the minimum $2\theta = 90°$.

$$\Delta(2\theta) = 2\Delta\theta = \left(\frac{\lambda}{2a}\right)^2 \frac{2}{\sin 2\theta} \tag{3}$$

In practice, the monochromator was selected before the "matching" procedure so that the parameters λ, $2\theta_M$ and β for the instrument were determined at first. A single crystal used as monochromator should have high neutron reflectivity, large enough mosaic spread β for high neutron intensity and a proper plane spacing d for the take-off angle $2\theta_M$ and the reflected neutron wavelength λ required by a specific spectrometer.

However, one has only a few choice since the growth of single crystals larger enough for the use is very difficult. For our texture diffractometer, a Cu (111) single crystal from the previous Four Circle Diffractometer (FCD) at Juelich Center for Neutron Science (JCNS) is used as the monochromator. Single crystals of Cu have relatively high neutron reflectivity and are often used as monochromators in neutron scattering instruments, but the mosaic spread β of this Cu (111) crystal is only about 6′, this value is too small for a texture instrument from the point of view of intensity, although lower β value benefits the resolution for angles off the minimum $A_{1/2}$ position. Fortunately it is a vertically bent monochromator, the intensity gain factor of 2-3can compensate nearly the intensity loss due to the too small β. A double focusing HOPG (002) or Si (311) monochromator will be equipped to this instrument in the future.

With the known inter planar spacing of Cu (111), d=2.0871Å, the neutron wavelength λ can be obtained from the Bragg equation $2d\sin\theta_M=\lambda$, if the take-off angle is given. Usually, the take-off angle $2\theta_M$ is put at about 2θ=90°, or even a larger angle, so that good match between the instrument and required resolution functions is obtained in a larger 2θ angle range. However, for our instrument the maximum $2\theta_M$ we could choose is restricted at about $2\theta_M=40°$ due to the limited geometric space, and thus the corresponding neutron wavelength is λ≈1.43Å. Although the use of this low take-off angle would depress the resolution of the instrument, it is still acceptable for texture measurement, whereas the corresponding neutron wave length λ is near the maximum of the incident neutron wave length spectrum, thus beneficial to the intensity [13-14].

With the given λ, β and $2\theta_M$, the required resolution functions were calculated for several typical metals to estimate the range of them and then a series of calculations were carried out for the instrument resolution function for different combinations of $\alpha_1$, $\alpha_2$ and $\alpha_3$.

Usually, in order to obtain high neutron intensity with relatively small influence on resolution, a large angle $\alpha_2$ is chosen for a diffractometer. In our case, the second collimator is needless since the natural collimation of ~30′ is already a reasonable value for $\alpha_2$. Therefore, all the calculations for the instrument resolution function were performed with $\alpha_2$ equal to 30′ fixed. By careful comparisons between the calculated instrument and required resolution functions, three different combinations of $\alpha_1$, $\alpha_3$ were finally adopted to meet requirements of different materials. The results are listed in table 1 and have been used in the construction of the texture diffractometer, and the corresponding instrument resolution functions are plotted in figure 2.

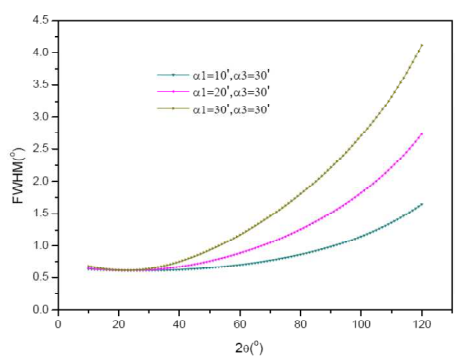

**Figure 2** The instrument resolution function for different combination of $\alpha_1$ and $\alpha_3$

Besides the monchromator and collimators, the sample table is another important component of the texture diffractometer, although it does not affect the intensity and resolution directly. For our instrument, the four circle mechanical device of the above mentioned Juelich FCD was modified and is reused as the sample table. With a four-circle device, the pole figure $P_{hkl}(\alpha,\beta)$ for a specified plane (hkl) can be obtained by rotation of the sample through $\chi$ (0°~90°) and $\varphi$ (0°~360°) angles with a certain step to align the various sample orientations along the (hkl) scattering vector, which is usually the bisector of the incident and the diffracted beam, and to record the diffracted intensities at each step respectively using a detector set at the 2θ position of the (hkl) reflection.

For the described pole figure measurement to be realized, the hardware and software for motion control and data acquisition were specially designed by the authors [15]. Shown in figure 3 is the hardware structure. The software was written in the Python language under the Linux system, and has been proved to be reliable by practical pole figure measurements

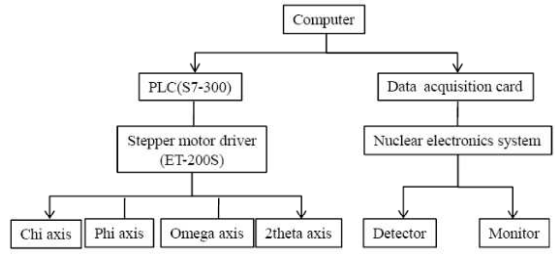

**Figure 3** Hardware structure of motion control and data acquisition

It is worth to mention that the $^3$He single detector used at present for the instrument will be replaced in the near future by a 200mm×200mm two dimensional position sensitive detector (PSD) supported by International Atomic Energy Agency. With this PSD, it is possible to measure several pole figures simultaneously with shorter time. Moreover, overlapping pole figures may be separated, this is especially important for the materials with line-rich diffraction patterns, such as intermetallic phases, ceramics and composites. The characteristics of this instrument are given in table1.

**Table 1** Characteristics of NTD at CARR

| | |
|---|---|
| Primary collimator | 10′, 20′, 30′ |
| Third collimator | 30′ |
| Monochromator | Vertical bent Cu(111) |
| Take-off angle | 42º |
| Wavelength | 1.48Å |
| Maximum Neutron flux at sample position | $5.6 \times 10^7$ n cm$^{-2}$s$^{-1}$ |
| Maximam beam cross section | 30×30 mm$^2$ |
| Detector | PSD  $^3$He tube |
| Distance monochromator-sample | 1800mm |
| Distance sample-detector | 500mm |

## 3  Experimental measurements

### 3.1 Diffraction pattern measurement for sample TiO$_2$

The diffraction pattern of a cylindrical TiO$_2$ sample with 20mm height and 10mm diameter was measured with α1=30′ to calibrate this instrument. The sample was mounted on a thin Al rod wrapped by a Cd foil and inserted into the center of the goniometer. The neutron beam cross section is 20mm

(width) × 30mm (height). 2θ-scan in the range of 25º to 72º was carried out at a step of 0.1º. The Rietveld profile technique was used to analyse the data with the program Fullprof, using the known structure of $TiO_2$.

Figure 4 shows the measured and calculated diffraction patterns of the $TiO_2$ sample. As can be seen, a good fit between them was achieved. The actual neutron wavelength λ and the corresponding take-off angle $2θ_M$ thus obtained, are listed in table 2. The real resolution curve of the instrument derived from the fit is plotted in fig 5, together with the designed one. Good agreement between them is also found.

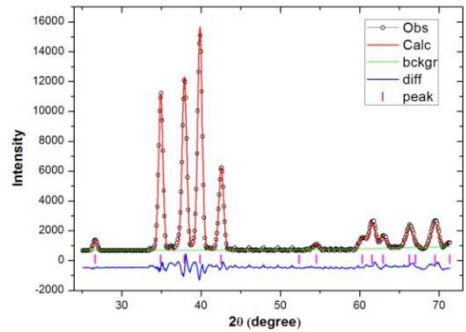

**Figure 4**  Neutron diffraction pattern of $TiO_2$ sample

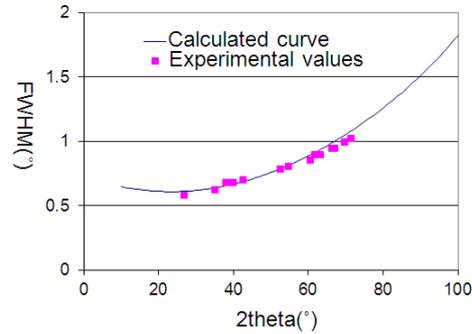

**Figure 5**  Resolution curve of the texture diffractometer at CARR

### 3.2 Neutron intensity measurement

The gold foil activation method was adopted to measure the neutron flux at the sample position. A circular gold foil with 19mm diameter was fixed at the center of the sample table, totally bathed in the neutron beam for about 6 hours at the reactor power of 10MW. Through measuring the actively of $^{198}Au$ produced by the reaction of $^{197}Au(n,γ)$ $^{198}Au$, the neutron flux was obtained. Corresponding to the reactor power of 60 MW, it is $5.6*10^7 n\ cm^{-2}s^{-1}$ at the sample position, and is also listed in table 2.

### 3.3 Pole figure measurements for warm-rolled Zircaloy-4 plate

The Pole figures of a Round-robin sample of warm-rolled Zircaloy-4 were measured to test the performance of this texture diffractometer. The texture measurementd for this sample have also been carried out on the neutron texture diffractometers of High-Pressure-Preferred Orientation (HIPPO) at LANSCE and Kowari at ANSTO. A cubic sample of $12\times12\times12$ mm$^3$ was prepared by spark cutting a warm-rolled Zircaloy-4 plate into small squares, and then gluing them together along the same rolling direction. The size of incident neutron beam was set as $25\times25$ mm$^2$ by the variable slit to ensure the sample bathing in the neutron beam.

The (10-10), (0002), (10-11) and (11-20) pole figures were measured respectively at the reactor power of 10 MW. The time taken by each pole figure is about 4 hours. The raw data were transferred into LaboTex format, then were used for quantitative ODF analysis. Shown in figure 6 are the (0001), (10-10) and (11-20) pole figures, recalculated by J.R. Santisteban from the obtained ODF together with those measured at HIPPO and Kowari. Good agreement is found both qualitatively and quantitatively between the different instruments [16-17]. The Kearns factors and texture indices were also calculated, the results are shown in table 2. All Kearns factors obtained on different instruments are within an uncertainty of ±0.02, which is lower than typical differences usually found for different batches, or between the start and end of pressure tubes [17-18].

**Table 2** Kearns factors and texture index for the Zircaloy-4 specimens

| Instrument | Transverse | Normal | Rolling | Sum | Text index |
|---|---|---|---|---|---|
| CARR-TD | 0.352 | 0.527 | 0.118 | 0.997 | 2.02 |
| HIPPO | 0.359 | 0.543 | 0.096 | 0.998 | 2.15 |
| Kowari | 0.357 | 0.551 | 0.090 | 0.998 | 2.32 |

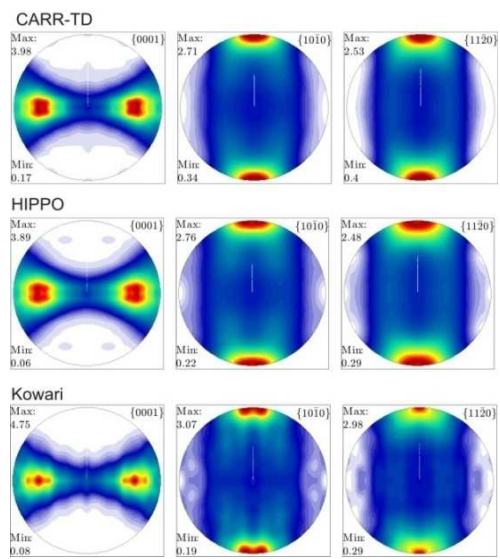

**Figure6** (0001), (10-10) and (11-20) pole figures recalculated from the ODF of warm-rolled Zircaloy-4 sample

## 4 Summary

The newly constructed neutron diffractometer at CARR is the first instrument for texture pole figure measurement by neutron in China. With relatively high neutron intensity and the performance, it has been proved to be efficient and suitable to pole figure measurements for simple structure materials in present stage. Up to now, a number of pole figures with satisfying quality have been obtained on this instrument for various research projects. A further improvement in its performance will be made in the near future by replacing the single $^3$He detector with a two dimensional PSD.

The NTD at CARR has been open to the user's community in China. Users from the county are welcome to apply this instrument for their research projects in dependently or in cooperation with us.

**Acknowledgements**

The author Meijuan LI thanks Dr. J.R. Santisteban for offering the Round-robin sample and comparing the texture results measured at different instruments, and also thanks Prof. Baisheng Zhang for helpful discussion. This work is supported by the National Nature Science Foundation of China (No. 11105231), International Atomic Energy Agency- TC program (No.CPR0012) and National Nature Science Foundation of China (No. 11205248).